\documentstyle[twoside,epsf]{article}

\input ibvs2.sty

\begin{document}

\IBVShead{5xxx}{xx Month 200x}

\IBVStitle{``MOST'' satellite photometry of Regulus}

\IBVSauth{Slavek Rucinski$^1$; Michael Gruberbauer$^2$;
David B. Guenther$^2$; Rainer Kuschnig$^3$; Jaymie M. Matthews$^4$;
Anthony F. J. Moffat$^5$; Jason F.\ Rowe$^6$;
Dimitar Sasselov$^7$; Werner W. Weiss$^3$}

\IBVSinst{Department of Astronomy and Astrophysics, University of Toronto. 
e-mail: rucinski@astro.utoronto.ca}

\IBVSinst{
Department of Astronomy and Physics, 
Saint Marys University, Halifax, N.S., Canada.
e-mail: (mgruberbauer,guenther)@ap.smu.ca}

\IBVSinst{Universit\"{a}t Wien, Institut f\"{u}r Astronomie, 
Wien, Austria.
e-mail: (rainer.kuschnig,werner.weiss)@univie.ac.at}

\IBVSinst{Department of Physics \& Astronomy, University of 
British Columbia, Vancouver, B.C., Canada.
e-mail: matthews@astro.ubc.ca}

\IBVSinst{D\'{e}partment de Physique, Universit\'{e} 
de Montr\'{e}al, Montr\'{e}al, QC, Canada.
e-mail: moffat@astro.umontreal.ca}

\IBVSinst{NASA Ames Research Center, Moffett Field, CA, USA.
e-mail: jasonfrowe@gmail.com}

\IBVSinst{Harvard-Smithsonian Center for Astrophysics, 
Cambridge, MA, USA.
e-mail: dsasselov@cfa.harvard.edu}

\IBVSobj{$\alpha$ Leo}
\IBVSobj{HD 87901}
\IBVSobj{HR 3982}

\IBVSabs{No eclipse has been found in 15 days of 
almost continuous photometry of $\alpha$~Leo with 
accuracy of about 0.0005 mag.}

\begintext

\setcounter{footnote}{7}

Regulus ($\alpha$~Leo) is a rapidly rotating, nearby B7V star which
has been suspected of small-scale variability and 
binarity for a long time, this in addition to the known 
binary K2V + M4V companion $3'$ away. But only recently 
Gies et al.\ (2008) have discovered that it is indeed a moderately
close binary with the orbital period $P = 40.11 \pm 0.02$ d. 
The discovery of radial velocity variations with the 
semi-amplitude of $K_1 = 7.7 \pm 0.3$ km~s$^{-1}$ 
was made in spite of the very strong broadening of the lines with
$V \sin i \simeq 320$ km~s$^{-1}$. The visible component moves 
radially by a distance only about twice its dimensions, 
$a_1 \sin i = 6.1 \pm 0.3\,R_\odot$. 
From the small value of the mass function and the assumed 
value for the mass of the primary $M_1 = 3.4 \pm 0.2 \, M_\odot$,
the authors derived $M_2 \ge 0.30 \pm 0.01 \, M_\odot$. 
Using various theoretical arguments on the evolution
of the Regulus binary system, Rappaport et al.\ (2009) argue
that indeed $M_2 = 0.30 \pm 0.02\,M_\odot$. 
The observed large value of
$V \sin i$ suggests that the axis of rotation and 
the orbital momentum may be positioned not far from 
the plane of the sky implying a possibility of eclipses. 

Chance and depth of eclipses depend on the size of the 
mutual orbit which -- in turn -- depends
on the mass ratio $q = M_2/M_1$. Thus, for the
case of Regulus' orbit, 
$(a_1+a_2) \sin i = 6.1\,(1+1/q)\,R_\odot$.
The secondary star cannot be large since it is 
spectrally undetectable; it can be a an M-type dwarf, 
a low-mass white dwarf or a low-mass helium star. 
As examples, for $i=90$ degrees and $q = 0.1$ or 0.2, 
the orbital dimensions 
would be $60\,R_\odot$ or $37\,R_\odot$. For
such a large orbit eclipses would take place only
within a small range of inclinations around $i=90$ deg.
For an infinitesimally small secondary component and 
the values of $q$ as above, the eclipses would occur for $i$
within $\pm 3$ or $\pm 5$ degrees away from the
edge-on orbital position; these margins would be
larger for physically larger stars as then chances of
a grazing eclipse would increase. For the two values of
$q$ and $i=90$ degrees, the 
eclipses could last up to 0.65 or up to 1.2 days,
respectively. 

\IBVSfig{10cm}{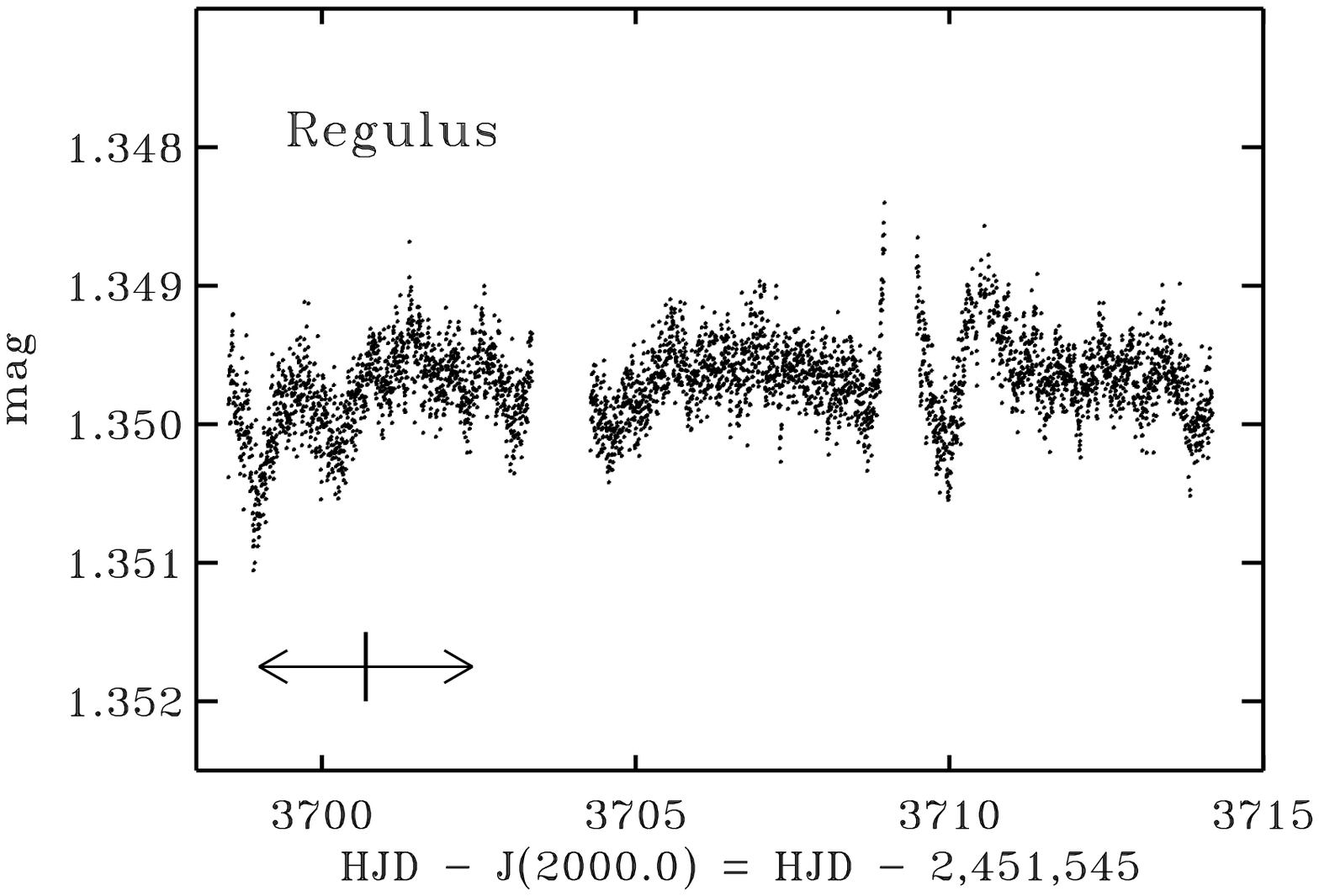}{MOST observations of Regulus
binned at 5 minute intervals. The vertical line and the
arrows mark the expected time of the conjunction with
the visible star behind.}

We attempted to discover eclipse transits 
of Regulus by the invisible companion using the MOST 
satellite\footnote{The MOST satellite is
a Canadian Space Agency mission, jointly operated 
by Dynacon Inc., the University of Toronto Institute of Aerospace 
Studies, and the University of British Columbia, with the assistance
of the University of Vienna.}. 
The optical system of the satellite consists
of a 15~cm reflecting telescope with a 
custom broad-band filter covering the spectral range of
380 -- 700~nm with the effective wavelength close
to the Johnson $V$ band.
The pre-launch characteristics of the mission are described
by Walker et al.\ (2003) and the initial post-launch performance
by Matthews et al.\ (2004). Since the failure of the
attitude control CCD in 2006, photometric observations are formed 
by adding short, typically one second exposures which are
needed for stabilization of the satellite. For Regulus, we used
the Fabry-lens mode with the star image spread within
$30 \times 30$ pixels. The temporal sampling after the
on-board addition was 30 sec. Because of the addition of the
read-out noise, the final mean standard error 
per single observation is a complex function
of the star brightness; it is expected to be at the
level of 0.25 mmag (milli-magnitude) 
for the brightness of Regulus (Kuschnig 2010, unpublished). 
It should be stressed that the satellite was designed
to be used for detection of {\it periodic signals\/} 
with time scales of minutes to hours and that 
long-term trends may happen and are sometimes hard
to characterize. Some of them can be removed by using 
stars simultaneously observed with the target or by following
satellite thermal and ambient magnetic field variations. 

The MOST observations of Regulus were done over 15 days, 
February 10 to March 4, 2010. The predicted time of the 
spectral inferior conjunction 
using the Gies et al.\ spectroscopic elements for $E=267$
elapsed epochs is: $T_0 + P/4 + E*P = $ 
HJD 2,455,245.70 or MOST time = 3700.70 (counted from J2000.0). 
Dr.\ Gies (private communication) estimated that this
time is uncertain by $\pm 1.7$ days.

The observations of Regulus are shown in Figure~1
after binning in 5 minute intervals, 
with the mean level adjusted to $V=+1.35$ which is the
normally observed magnitude of the star; note
that -- as common for brightest stars -- the scatter
in the literature values of $V$ is large reaching
$\pm 0.02$ mag. We show the
whole data well beyond the predicted moment of the
eclipse to illustrate that the small depressions observed
at the predicted conjunction time may be spurious
and cannot be interpreted as an eclipse.
Similar fluctuations which reach 0.5 mmag of the mean signal
and are present throughout the duration of the whole run
could not be eliminated using any known
instrumental effects. This is best visible around 
occurrences of two breaks of the sequence for 0.9 and 0.4
days which were caused by the telescope solar-door problem
and an interruption to monitor a super-Earth transit. 

The residual variability seen in Regulus cannot 
be unambiguously interpreted as coming from the 
star, since the background measurements and telemetry 
show variability on similar time scales. Frequency 
analysis of the data and the telemetry did not reveal 
significant, periodic, coherent variations that would be
clearly unique to Regulus at the amplitude level of
about $<0.07$ mmag ($7 \times 10^{-5}$ mean signal). 
Although the frequency range 0.3 to 3 cycles per day
may require further investigation, at this point
we have no convincing evidence 
for variations related to the rotation of Regulus 
at a frequency of about 1.7 cycles per day.

Summarizing: MOST observations did not lead to detection of 
any obvious eclipse deeper than about 0.5 mmag at the
predicted moment of the spectroscopic inferior conjunction. 
For an orbit inclined by $i > 87$ degrees
this excludes a red dwarf with $M_2 \simeq 0.3\,M_\odot$
as a companion because such a star would produce an
eclipse up to 8 mmag deep. 
However, a low-mass white dwarf or a helium star 
-- which according to Rapport et al.\
-- are more likely candidates for a companion of Regulus 
would be undetectable by MOST:
With their expected radius $R \simeq 0.02 - 0.06\,R_\odot$,
the eclipse would be only 0.04 to 0.3 mmag deep.

\references

Gies, G. R., Dieterich, S., Richardson, N. D., Riedel, A. R., 
    Team, B. L., McAlister, H. A., Bagnuolo Jr., W. G., 
    Grundstrom, E. D., Stefl, S., Rivinius, T., Baade, D.,
    2008, {\it ApJ\/}, {\bf 682}, L117

Matthews J.M., Kusching R., Guenther D.B., Walker G.A.H.,
    Moffat A.F.J., Rucinski S.M., Sasselov D., Weiss W.W.
    2004, {\it Nature\/}, {\bf 430}, 51

Rappaport, S., Podsiadlowski, Ph., Horev, I.
    2009, {\it ApJ\/}, 698, 666

Walker G., Matthews J., Kuschnig R., Johnson R., Rucinski S.,
    Pazder J., Burley G., Walker A., Skaret K., Zee R., Grocott S.,
    Carroll K., Sinclair P., Sturgeon D., Harron J. 
    2003, {\it PASP\/}, {\bf 115}, 1023

\endreferences 

\end{document}